# Ultrafast lattice deformations studied by means of time-resolved electron and x-ray diffraction

Runze Li[1*], Kyle Sundqvist[2], Jie Chen[3], H. E. Elsayed-Ali[4] and Peter M. Rentzepis[1*]

[1]*Department of Electrical and Computer Engineering, Texas A&M University, College Station, TX 77843, USA*
[2]*Department of Physics, San Diego State University, San Diego, CA 92182, USA*
[3]*Center for Ultrafast Science and Technology, Key Laboratory for Laser Plasmas (Ministry of Education), School of Physics and Astronomy, Collaborative Innovation Center of IFSA (CICIFSA), Shanghai Jiao Tong University, Shanghai 200240, China*
[4]*Department of Electrical and Computer Engineering, Old Dominion University, Norfolk, VA 23529, USA*
* *Email:Runze@tamu.edu; prentzepis@tamu.edu*

## Abstract

Ultrafast lattice deformation of tens to hundreds of nanometer thick metallic crystals, after femtosecond laser excitation, was measured directly using 8.04 keV subpicosecond x-ray and 59 keV femtosecond electron pulses. Coherent phonons were generated in both single crystal and polycrystalline films. Lattice compression was observed within the first few picoseconds after laser irradiation in single crystal aluminum, which was attributed to the generation of a blast force and the propagation of elastic waves. The different time scale of lattice heating for tens and hundreds nanometer thick films are clearly distinguished by electron and x-ray pulse diffraction. The electron and lattice heating due to ultrafast deposition of photon energy was numerically simulated using the two-temperature model (TTM) and the results agreed with experimental observations. The ultrafast heating described by TTM was also discussed from an electrical circuit perspective, which may provide new insights on the possible connection between thermal and electrical processes. This study demonstrates that the combination of two complimentary ultrafast time-resolved methods, ultrafast x-ray and electron diffraction will provide a panoramic picture of the transient atomic motions and structure in crystals.

## Introduction

Probing dynamics with a sub-angstrom structural and femtosecond temporal resolution at the atomic level is the key to track the pathways and intermediate states in chemical reactions and understand the functions of materials at a fundamental level. Because the atomic vibrational time in lattice is typically on the order of a few femtoseconds, there are no mechanical or electronic means that are sufficiently fast to capture such ultrafast processes. However, following the time-resolved studies of pulsed-photon-induced reactions in the early 1950s [1], the pump-probe method was developed, which demonstrated for the first time, a picosecond temporal resolution for chemical reactions [2, 3]. Several years later, after the introduction of femtosecond laser pulses [4], tracing the dissociation and formation of chemical bonds, that occur within femtosecond time interval, became possible [5]. Among the various pump/probe combinations, optical pump/optical probe [6, 7], optical pump/electron probe [8-12], and optical pump/x-ray probe [13-16] are the three most widely used experimental means. Purely optical methods, such as transient spectroscopy[17] or transient reflectivity measurement [18], usually provide information on the relaxation of the electron systems after laser excitation, while the dynamics concerning the atomic structure are obtained indirectly through theoretical modeling. Ultrafast time-resolved diffraction methods that employ x-ray [19-21]  and/or electron pulses [22-27] , however, provide direct information on the structural dynamics as they probe directly the atoms and/or inner shell electrons. The femtosecond and picosecond x-ray pulses, can be generated by large national facilities, free electron lasers [28-30] and small compact systems that are based on laser-plasma interaction [31] and are suitable for university size laboratories. The femtosecond electron pulses are usually generated by photo-electron guns and accelerated by table-top DC electric fields [32, 33] or RF fields [27, 34, 35] to tens of keV or a few MeV, respectively. In time-resolved electron diffraction studies, the presence of tens of kV/m transient electric fields on the metallic or semiconductor sample [36-38] may make the interpretation of electron diffraction data difficult. However, as charged particles, electron pulses are suitable for diagnosing fast-evolving plasmas or warm dense matter that involves transient electro-magnetic fields [39-42]. In contrast, x-ray pulses are insensitive to transient electric fields and therefore provide a "clean" signal of ultrafast structural information of the crystalline samples. The limited yield and the flux instability of hard x-ray photons generated by the laser-plasma sources, however, make it rather difficult to obtain a high signal-to-noise ratio for diffraction intensities. In addition, the optical skin depth of metals is typically on the order of a few nanometers, which is close to the penetration depth of keV electron beam while much shorter than that of the x-ray penetration depth into the bulk of the sample. Therefore, for tens of nanometer thick samples, where homogeneous laser excitation is expected, it is more accurate and convenient to use electron pulses, while for hundreds nanometer thick samples, it is more appropriate to use x-ray pulses that probe not only the surface but also the bulk that is heated due to the laser energy deposited onto the surface layer and transferred to the interior of the crystals. Therefore, using results from both time-resolved ultrafast electron and x-ray diffraction, which are two highly complementary methods, may provide a more detailed description of the transient structural dynamics of crystals, than either methods can provide alone [43].

In this paper, the transient atomic motions of aluminum crystals, illuminated with femtosecond laser pulses, were recorded in real time with the combination of x-ray and electron probes. The ultrafast heating of electron and lattice subsystems were simulated using the two-temperature model (TTM), which agrees with experimental observations. Meanwhile, an electronic circuit representation of the laser energy deposition process into the crystal is also described, which provides a new perspective for ultrafast heat transport processes within crystals.

**Experimental Method**

The methods used in our experiments are demonstrated in Figure 1. The ultrafast time-resolved x-ray diffraction experiments were performed on a table-top system that generates hard x-ray pulses through laser/plasma interaction [44]. The 800 nm, 100 fs, 100 mJ laser pulse emitted from a 10 Hz Ti:sapphire laser was split into two parts. 80% of the laser energy is focused onto a 0.5 mm diameter copper wire that is continuously moving in a vacuum chamber to generate 8.04 keV Cu K$\alpha$ sub-picosecond x-ray pluses. The upper limit of the x-ray pulse duration is estimated to be 0.6 ps [45]. The x-ray pulses were collimated by a 200 $\mu$m slit and impinged onto a 150 nm thick single Al (111) sample in a reflection diffraction configuration with a Bragg diffraction angle of 18.9 degree. The remaining 20% of the laser energy is frequency doubled to 400 nm and used to excite the sample. Before focusing onto the sample, the 400 nm excitation laser pulse is directed to a linear translation stage that precisely controls the arrival time between the probe x-ray pulses and the pump laser pulses. The Al (111) single crystal used in our experiments was grown on Mica substrate at a base temperature of 150 °C.

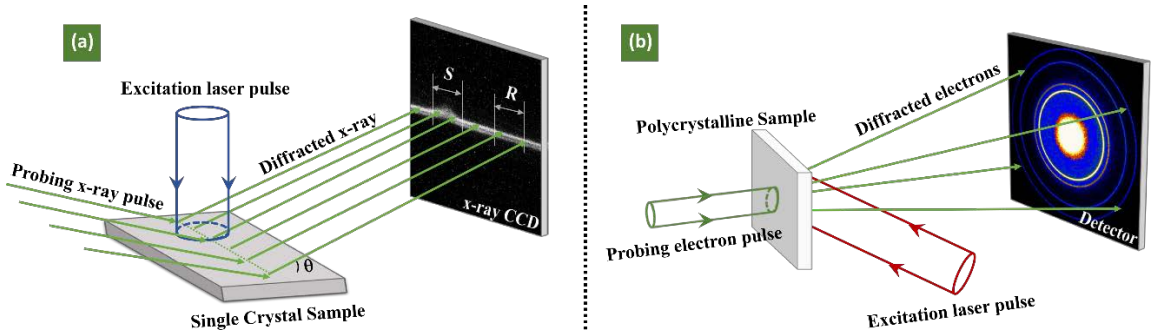

Figure 1. Schematic diagram of the ultrafast time-resolved x-ray and electron diffraction methods. (a) The diffraction from a single crystal sample investigated by sub-picosecond x-ray pulses. The x-ray diffraction signal originated from the laser excitation area of the Al (111) crystal was marked "S" on the detector, while "R" denoted the reference without laser irradiation. (b) The electron diffraction for polycrystalline Al sample investigated by femtosecond electron pulses. The angle between the front-illumination laser and back-probing electron beams is ~10° to minimize the geometric mismatch and improve temporal resolution.

The ultrafast time-resolved electron diffraction experiments were performed on a table-top system that is capable of delivering sub-picosecond, 59 keV, electron pulses [37]. The 800 nm, 70 fs, 1 mJ laser pulse, emitted by a 1 kHz Ti:sapphire laser system are split into two parts by a beam splitter. 40% of the laser energy is frequency tripled to 266.7 nm and directed to the silver photocathode, using a back illumination, to generate electrons, which were accelerated to 59 keV by a DC electrical field and focused by a magnetic lens onto the sample to form a transmission diffraction configuration in the ultra-high vacuum chamber ($< 1 \times 10^{-9}$ Torr). The remaining 60% of the laser energy, which is attenuated and used to excite the polycrystalline aluminum sample, is directed to a translation stage that can control precisely the arrival time between the pump laser pulse and the probe electron pulse at the sample. The temporal-resolution of the ultrafast time-resolved electron diffraction system is estimated to be 0.5 ps. The 10 ~ 20 nm thick polycrystalline Al films used in this study were obtained as follows: first they were deposited onto freshly cleaved NaCl single crystal surfaces, then immersed into deionized water and subsequently transferred to transmission electron microscope (TEM) grids as freestanding films.

The x-ray diffraction patterns were recorded directly by a 2D x-ray charge-coupled device (CCD) while the electron diffraction patterns were converted into optical signals on a phosphor screen, amplified by an image intensifier and eventually recorded by a 2D optical CCD. The time-dependent changes in the diffraction peaks were analyzed and correlated with transient atomic motions. According to the first-order Bragg diffraction equation, $2d\sin\theta = \lambda$, where $d$ is the lattice plane distance, $\theta$ is the diffraction angle and $\lambda$ is the wavelength of probing electrons of x-ray photons, one can obtain the following relation: $\Delta d/d = -\Delta\theta/\tan\theta$, which directly connects the relative change of the lattice plane distance with the change of diffraction angles, obtained from the experimentally recorded diffraction patterns. For keV or MeV electron diffraction, because the diffraction angle is typically small (less than 1º), the small angle approximation may be applied and therefore one could use $\Delta d/d = -\Delta\theta/\tan\theta \approx -\Delta\theta/\theta$.

**Results and Discussion**

**1. Transient structural changes of single crystal and polycrystalline aluminum films**

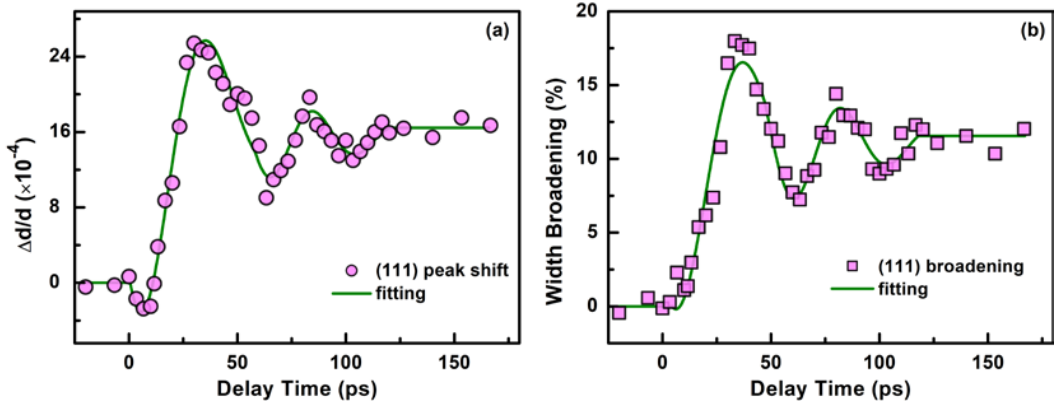

Figure 2. Time-dependent relative change of (a) diffraction peak position and (b) diffraction line width boarding of a 150 nm thick single crystal Al (111) film interrogated by sub-picosecond x-ray pulses in a reflection diffraction configuration. The excitation laser fluence is 18.5 mJ/cm².

The transient structural changes of the 150 nm thick Al (111) single crystal after illumination with 400 nm femtosecond laser pulses was probed by 8.04 keV sub-picosecond x-ray pulses. The diffraction peak shift, which represents the relative changes in the lattice plane distance, and the broadening of the diffraction line width as a function of time is shown in Figure 2. The data show that, the relative change of the lattice plane distance is negative during the first few picoseconds after laser excitation, which reveals that contraction takes place and the lattice plane distance becomes shorter. Following the contraction, damping oscillations of the lattice plane were developed, accompanying a lattice plane expansion. The new equilibrium position of the lattice plane distance was formed ~120 ps after illumination with the femtosecond pulse(s). The initial contraction of the lattice plane distance is due to the formation of a compressive wave that was initiated on the surface layer of the crystal as a result of a blast force [46-49]. Because the x-ray pulse penetrates and probes the entire 150 nm thick crystal, the contraction formed within the top few lattice layers account for less than 10% of the overall diffraction signal.

Therefore, the contraction observed during the first few picoseconds is small, ~8%, compared with the observed expansion of the lattice. Such contraction was also observed previously in Au single crystals [50]. After the contraction stage, the laser energy initially deposited within the skin depth of the crystal propagates through the bulk and heats the entire bulk of the crystal through electron-phonon and phonon-phonon interaction. The elevated crystal temperature is reflected by the increased shift of the diffraction line peaks and the width. The increase of the lattice plane distance is accompanied by several damping oscillations, which indicate that coherent phonons were generated. Such lattice damping oscillations have been observed in a number of time-resolved diffraction studies [20, 51, 52]. These oscillations represent that the vibration of the lattice and the damping is accompanied by the dissipation of energy into the surroundings or the conversion of kinetic energy into heat. The mechanism of acoustic oscillations may be explained theoretically using the thermal strain-stress model or the Fermi-Pasta-Ulam anharmonic chain model [53-55]. The optical phonons are not addressed in this study. The oscillation is generally simplified by considering it as the formation of a one-dimensional standing wave between the surface and crystal substrate. Therefore, the oscillation period, $T$, may be calculated using the sound velocity of the crystal: $T = 2L/v$, where $L = 150$ nm is the film thickness and the sound velocity in solid aluminum $v = $ 6420 m/s [56]. The oscillation period was calculated to be ~47 ps, which is in agreement with our experimentally observed value of ~ 41 ps. The diffraction peak shift remains unchanged after the damping oscillation for ~ 180 ps, which is the longest delay time in our experiments.

The transient atomic motions measured by sub-picosecond x-ray pulse experiments were also compared with those obtained through ultrafast time-resolved electron diffraction experiments. To avoid the effects of transient electric fields, on the shift of the electron diffraction peak [37], a transmission diffraction configuration was employed in the ultrafast electron diffraction experiments. Polycrystalline aluminum films of three different thicknesses, 10 nm, 15 nm and 20 nm were interrogated by 59 keV, sub-picosecond electron pulses. The relative changes of the (311) lattice plane distance for those films are shown in Figure 3 and the theoretical and experimental periods versus crystal thickness are listed and compared in Table 1. It is worth mentioning that, because the 10 nm thick film contains only a few layers of lattice, it might not be a continuous polycrystalline film and is most likely built up with flat connected islands. It is clear that, the lattice expansion was observed immediately after the laser excitation at a fluence of ~2.1 mJ/cm$^2$. The maximum expansion of the lattice plane distance is achieved within a few picoseconds after laser excitation and the new equilibrium state is established around tens of picoseconds, which is much shorter than the equilibrium time observed in x-ray diffraction experiments presented above. This is because the film thickness in electron diffraction experiments is comparable or close to the penetration depth of laser pulses. Therefore, the entire crystal is expected to be heated evenly within the first few picoseconds and contribute homogeneously to the electron diffraction pattern. In the x-ray diffraction case, however, the excitation laser energy is also absorbed within the skin depth, while the x-ray pulse probes the entire crystal, which in this case is about twenty times longer than the skin depth. Therefore, a much longer time is required to transfer the heat to the entire probing depth of the sample which travels with sound velocity. This process requires tens or hundreds of picoseconds depending on the thickness of the crystal.

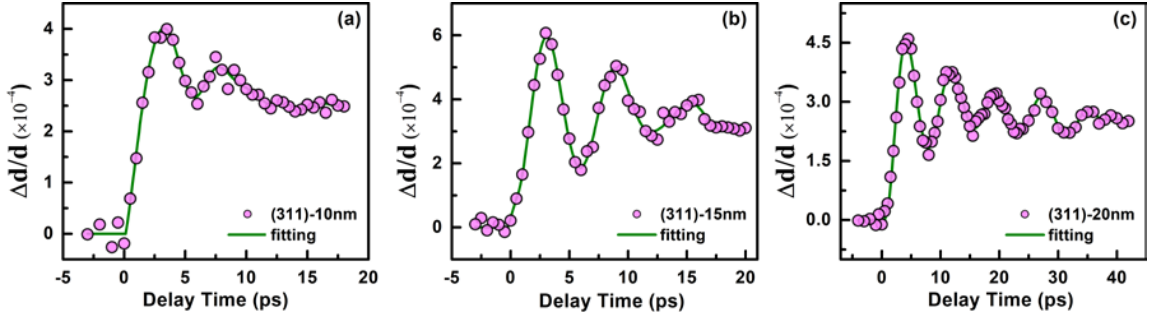

Figure 3: Time-dependent relative change of the (311) lattice plane distance of (a) 10 nm, (b) 15 nm and (c) 20 nm thick polycrystalline aluminum freestanding films interrogated by femtosecond electron pulses in a transmission diffraction configuration. The laser pump fluence is 2.1 mJ/cm² on each sample.

Table 1: Oscillation periods of Al samples with three different thicknesses. The theoretical period was calculated using the one-dimensional standing wave model.

| Sample Thickness (nm) | 10±5 | 15±5 | 20±5 |
|---|---|---|---|
| Theoretical Oscillation Period (ps) | 3.1±1.6 | 4.7±1.6 | 6.2±1.6 |
| Experimental Oscillation Period (ps) | 3.8 | 6.0 | 6.4 |
| Sample thickness inferred from experimental period (nm) | 12 | 19 | 21 |

## 2. Electron and lattice heating described by the Two-temperature Model (TTM)

Upon femtosecond laser irradiation, the photon energy is initially deposited into the free electrons within the skin depth of the 150 nm thick Al (111) crystal within the duration of the femtosecond laser pulse. Tens or hundreds of femtoseconds after the initial stage, the entire electron system reaches equilibrium through electron-electron interaction. Because the heat capacity of the electrons, $C_e$, is typically orders of magnitude smaller than that of the lattice heat capacity, $C_l$, the electron temperature may reach tens of thousands of Kelvin while the lattice remains cold, at room temperature. Subsequently, the heat transport between electrons and lattice occurs by means of electron-phonon interaction. The rate of energy exchange between the electron and lattice system is described by the electron-phonon coupling，$g$，and the temperature equilibration between the electron and lattice subsystems usually takes place within picoseconds through electron-phonon, phonon-phonon, and phonon-lattice interactions. Heat transport within metals illuminated with ultrashort laser pulses is generally described by the well-established two-temperature model, in which the temperature evolution of the electron and lattice subsystems are described by two coupled heat transport equations. Because the area probed by the x-ray or electron pulses is several times smaller than that of the diameter of the laser excitation area, the three-dimensional laser-matter interaction could be reduced to a one-dimensional representation along the sample thickness direction. The electron and lattice heating of the 150 nm aluminum crystal, in this study, may therefore be described by the following two-temperature model (TTM) [9, 57, 58]:

$$C_e \frac{\partial}{\partial t} T_e(z,t) = -\frac{\partial}{\partial z} P(z,t) - g\left[T_e(z,t) - T_L(z,t)\right] + S(z,t) \tag{1}$$

$$C_L \frac{\partial}{\partial t} T_L(z,t) = g\left[T_e(z,t) - T_L(z,t)\right] \tag{2}$$

$$P(z,t) = -\kappa_e \frac{\partial}{\partial z} T_e(z,t) \tag{3}$$

where $T_e(z,t)$ and $T_L(z,t)$ are the electron and lattice temperatures at the depth $z$ under the aluminum sample surface at a certain time $t$, respectively. The excitation laser pulse is described by $S(z,t)$ which has a Gaussian profile and $k_e$ is the electron thermal conductivity. Ballistic electron motion is not considered, in this cause, because it contributes insignificantly to the heating of the film. Under low excitation fluences, where the electron temperature may increase by only a few hundreds or thousands Kelvin, it is appropriate to assume that the electron-phonon coupling, $g$, the electron heat capacity and thermal conductivity, $C_e$ and $k_e$, are constants or linearly dependent on the electron temperature. For electron temperatures that may increase by more than tens of thousands Kelvin, more sophisticated modeling maybe required. There has been a large number of literature papers concerning the application of TTM to time-resolved diffraction studies [25, 59, 60]. Here, we use the one that is applicable to both low and higher laser excitation fluences [52, 61]. In the following simulation, the Drude model is used for electron-electron and electron-phonon scattering, therefore, the electron thermal conductivity (W/mK) is given by [62-64]:

$$\kappa_e = \frac{a_0 T_e}{a_1 T_e^2 + a_2 T_L} \tag{4}$$

where $a_0 = 1.08 \times 10^{14}$, $a_1 = 5.2 \times 10^6$, and $a_2 = 4.61 \times 10^{11}$. The electron-phonon coupling, $g$ (W/m$^3$K), and electron heat capacity, $C_e$ (J/m$^3$K), are provided by the fifth-order Padé approximations:

$$g = 1 \times 10^{17} \frac{\sum_{n=0}^{5} A_1(n) \left( \frac{T_e}{10^4} \right)^n}{1 + \sum_{n=1}^{5} A_2(n) \left( \frac{T_e}{10^4} \right)^n} \tag{5}$$

$$C_e = 1 \times 10^5 \frac{\sum_{n=0}^{5} B_1(n) \left( \frac{T_e}{10^4} \right)^n}{1 + \sum_{n=1}^{5} B_2(n) \left( \frac{T_e}{10^4} \right)^n} \tag{6}$$

where $A_1(0) = 2.4897$, $A_1(1) = -3.5228$, $A_1(2) = 99.2859$, $A_1(3) = -125.4458$, $A_1(4) = 116.2749$, $A_1(5) = -19.5488$ $A_2(1) = -0.6301$, $A_2(2) = 24.2843$, $A_2(3) = -29.5203$, $A_2(4) = 28.8661$ and $A_2(5) = -4.8065$; $B_1(0) = 0.0348$, $B_1(1) = 5.7659$, $B_1(2) = 51.9519$, $B_1(3) = -16.9218$, $B_1(4) = 121.2319$ and $B_1(5) = -17.4062$; $B_2(1) = 2.0446$, $B_2(2) = 7.087$, $B_2(3) = 3.3556$, $B_2(4) = 2.0852$ and $B_2(5) = -0.3873$.

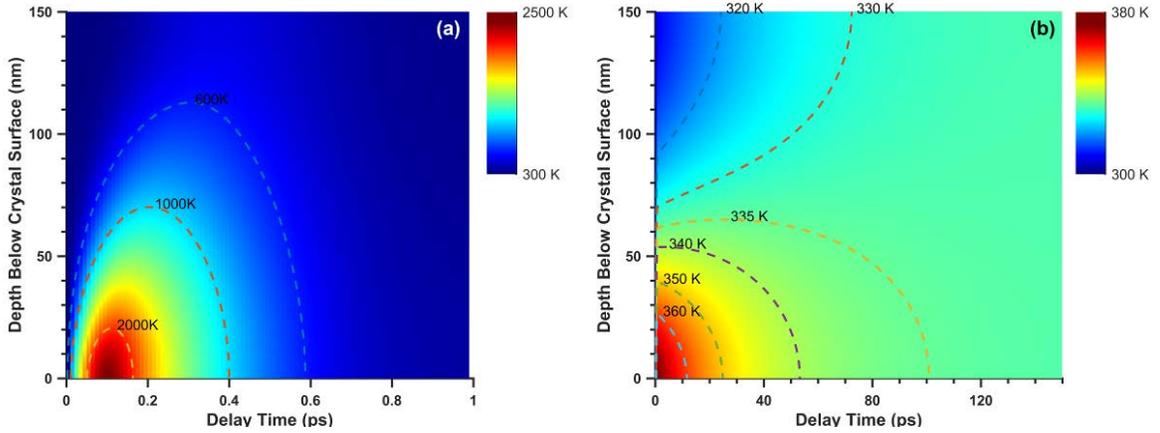

Figure 4: Two-Temperature Model simulation of the 150 nm thick aluminum crystal illuminated by 18.5 mJ/cm$^2$ femtosecond laser pulse. (a) spatial-temporal distribution of the electron temperature; (b) spatial-temporal distribution of the lattice temperature.

The spatial-temporal distributions of both electron and lattice temperatures, of the 150 nm thick Al (111) crystal, after femtosecond laser excitation are simulated by the theoretical model discussed above and plotted in Figure 4. It is found that the highest electron temperature of 2500 K can be reached within 100 fs, which is essentially the pulse width of the femtosecond laser pulse(s) used in our experiments. The thermalization, of the electron system within the 150 nm thick Al crystal, occurring in less than 1 ps, is accompanied by the energy transfer from the electron system to the lattice system through electron-phonon coupling. The lattice temperature may reach as high as 380 K within the first few picoseconds, after laser irradiation, and then equilibrate with the bulk of the crystal through thermal diffusion and phonon-phonon interaction. Our numerical simulation suggested that the largest non-equilibrium temperature occurs during the first 30 picoseconds, which agrees with the experimental results shown in Figure 2 where the lattice expansion kept increasing for the first tens of picoseconds after laser excitation. The thermal equilibrium state of the entire 150 nm crystal may be established within 100 ps. However, it should be noted that, because this model describes only the temperature evolution and does not simulate the motion of each individual atom, the lattice oscillations observed in the experiments discussed here are not represented completely in the simulation.

## 3. An electrical analog of the two-temperature model using transmission lines and circuit theory

To gain insight into the basic dynamics of the two temperature model, we also analyze this system using transient waves on a 1D transmission line. Here we treat wave properties in the linear limit, and while specific heat and thermal conductivity are known to vary with electron temperature, our transmission model is not concerned with temperature-dependent non-linearity. We consider first a conductive model of a single propagating temperature wave and neglect boundary conditions. The specific heat per unit length relates the changes in the stored heat, in a given mass to temperature change, as an analog to the telegrapher's electrical equation that relates the voltage time derivative to electrical current gradient.

$$C\frac{\partial}{\partial t}T(z,t) = -\frac{\partial}{\partial z}P(z,t) \qquad (7)$$

$$P(z,t) = -\kappa \frac{\partial}{\partial t} T(z,t) \qquad (8)$$

The heat conduction analog to Faraday's Law and to the 2nd telegrapher's equation, Eq. 8, relate the power per area to the temperature gradient. The equivalent, distributed circuit for heat conduction, is shown in Figure 5(a) over an infinitesimal length $\Delta z$. We may now interpret the three linearly independent equations of the TTM, Eq. 1-3, as the equivalent to the two telegrapher's equations of the simple, single-temperature case. Figure 5(b) depicts how the TTM may be represented as the transmission line model. The analog electrical and thermal variables are listed in Table 2. Owing to the fact that lateral heat conduction in the TTM model occurs only in the electrical system, with no lattice conduction counterpart, we employ a single transmission line.

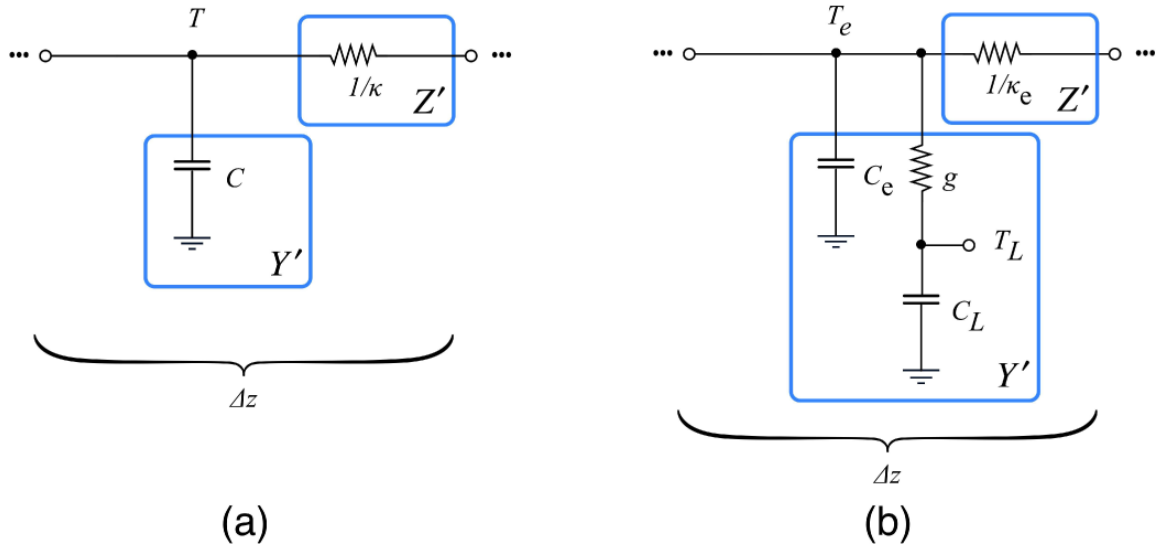

(a)                    (b)

Figure 5: Equivalent transmission-line models, over infinitesimal distance $\Delta z$, of (a) a simple heat equation with a heat capacitance-per-length and conduction-per-length, and of (b) the full TTM, as depicted in analogs to "telegrapher's equations" of Eq. 1-3, which contain both a distributed electron temperature and lattice temperature.

Table 2: Electrical and thermal analog variables

| Electrical | V | I | R | L | G | C |
|---|---|---|---|---|---|---|
| Thermal | $T$ | $P$ | $1/\kappa_e$ | -- | $g$ | $C$ |

In the frequency domain, this system may be described by a Laplace formalism, where the lattice temperature is now considered as a "voltage divided" electron temperature $T_l = gT_e/(g + sC_l)$. The thermal "impedance per unit length" is related to the thermal conduction: $Z^{'} = 1/\kappa_e$. The thermal "admittance per unit length" is: $Y^{'} = sC_e + sC_l/(1 + sC_l/g)$, and the characteristic impedance is therefore: $Z_0 = \sqrt{Z^{'}}/\sqrt{Y^{'}} = \sqrt{(1 + g/C_l)/C_e\kappa_e s\left[s + g\left(1/C_e + 1/C_l\right)\right]}$. The complex propagation constant is also therefore $\gamma = \sqrt{Z^{'}}\sqrt{Y^{'}} = \sqrt{(C_e/\kappa_e)\cdot s\cdot\left[s + g\left(1/C_e + 1/C_l\right)\right]/(s + g/C_l)}$.

Using $s = j\omega$, we can find attenuation coefficient and wavenumber:

$$\alpha = \sqrt{\frac{\omega}{\kappa_e}}\left(C_e^2 + \frac{C_l\left(2C_e + C_l\right)}{1 + C_l^2\omega^2/g^2}\right)^{1/4}\cos\left[\frac{1}{2}\arg\left(j + \frac{C_l}{C_e\left(C_l\omega/g - j\right)}\right)\right]$$

$$\beta = \sqrt{\frac{\omega}{\kappa_e}}\left(C_e^2 + \frac{C_l\left(2C_e + C_l\right)}{1 + C_l^2\omega^2/g^2}\right)^{1/4}\sin\left[\frac{1}{2}\arg\left(j + \frac{C_l}{C_e\left(C_l\omega/g - j\right)}\right)\right]$$

The complex propagation constant varies greatly as a function of frequency, which indicates that propagation in this system is highly dispersive. The attenuation and wavenumber, both, depend on the admittance per unit length, $Y^{'}$, and therefore it is important to analyze their behavior along with $Y^{'}$ as a function of frequency. At the lowest frequencies, $Y'$ is dominated by the heat capacitances and appears approximately as $j\omega(C_l + C_e)$. Under this condition, we consider the analogy to the condition of propagation in an electrical conductor dominated by skin-depth effects (i.e., $\alpha = \beta$) [65, 66]. Here, $\alpha = \beta$ both vary as $\sqrt{\omega}$. It is important to point out that the condition $\alpha = \beta$ results in substantially damped propagation, with loss occurring at the attenuation rate of 55 dB per unit wavelength (figure 6b) [67]. At frequencies spanning $g/C_l < \omega < g\left(1/C_l + 1/C_e\right)$, the $Y'$ is dominated by the electron-phonon dissipative term, $g$.

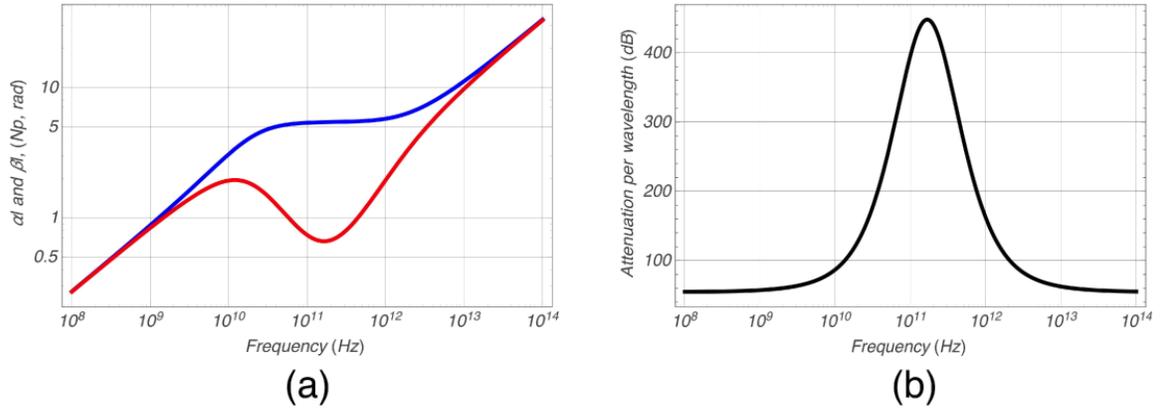

(a)                                                        (b)

Figure 6: The frequency dependence in the TTM for (room-temperature) bulk aluminum. (a) the attenuation ($\alpha l$) and wavenumber ($\beta l$) for a thickness of $l = 150$ nm. (b) the attenuation per wavelength in dB.

Since g resembles essentially electrical conductance, $\alpha$ becomes frequency-independent and dominates over an exceedingly small $\beta$. The transition to this regime provides also for a negative group velocity ($\partial\omega/\partial\beta < 0$). In this range, the attenuation rate is even more extreme - exceeding even 400 dB per wavelength. At larger frequencies, $\omega > g\left(1/C_l + 1/C_e\right)$, $Y^{'}$ is dominated only the electronic heat capacitance, ($Y^{'} = j\omega C_e$) and the behavior re-emerges that $\alpha = \beta$ with both proportionate to $\sqrt{\omega}$.

**Conclusion**

Using sub-picosecond 8.04 keV x-ray pulses and 59 keV femtosecond electron pulses as probes, we systematically probed and recorded the transient atomic motions of single crystal and polycrystalline aluminum films irradiated with femtosecond laser pulses, respectively. Coherent phonons were generated and observed in aluminum films of different thicknesses, and the experimentally observed oscillation periods agree with the values suggested by one-dimensional standing wave model. Lattice contraction of the single crystal was detected, which indicated the generation of blast force due to the non-equilibrium electron pressure. The different time scales for heating tens and hundreds of nanometer thick metallic lattices are clearly distinguished by electron and x-ray pulse experiments. The electron and lattice heating processes were numerically simulated using the two-temperature model and the results agree with experimental observations. An electrical circuit representation of the two-temperature model is described, which provide new insights between the transient thermal and electrical processes. The combination of these two methods, in a single experimental frame, has demonstrated in this study and provided a powerful means for obtaining detailed information of the transient atomic motions in crystals of various thicknesses.

## Acknowledgements:

We gratefully acknowledge the partial support by The Welch Foundation (Grant No.: 1501928) and Texas A&M University TEES funds. Runze Li thanks the helpful discussion with Prof. Jie Zhang. Jie Chen thanks the support by the National Natural Science Foundation of China under Grant Nos. 11574196, 61222509, and 11421064. H. E. Elsayed-Ali acknowledges support by National Science Foundation grant DMR-170817.